# GOTO Rankings Considered Helpful[1]

Emery Berger - University of Massachusetts Amherst, USA - emery@cs.umass.edu
Stephen M. Blackburn - Australian National University, Australia - steve.blackburn@anu.edu.au
Carla Brodley - Northeastern University, USA - c.brodley@northeastern.edu
H. V. Jagadish - University of Michigan, USA - jag@umich.edu
Kathryn S. McKinley - Google, USA - ksmckinley@google.com
Mario A. Nascimento - University of Alberta, Canada - mario.nascimento@ualberta.ca
Minjeong Shin - Australian National University, Australia - minjeong.shin@anu.edu.au
Lexing Xie - Australian National University, Australia - lexing.xie@anu.edu.au

Rankings are a fact of life. Whether or not one likes them (a recent CACM editorial argued that we should eschew rankings altogether [Vardi 2016]), they exist and are influential. Within academia, and in computer science in particular, rankings not only capture our attention but also widely influence people who have a limited understanding of computing science research, including prospective students, university administrators, and policy-makers. In short, rankings matter.

Today, academic departments are mostly ranked by for-profit enterprises. The people doing the ranking are not computer scientists, and typically have very little understanding of our field. For example, *US News and World Report*, in ranking PhD programs in sub-areas of Computer Science inaccurately describes the characteristics of research in the area of "Programming Language" [sic] (Fig. 1):

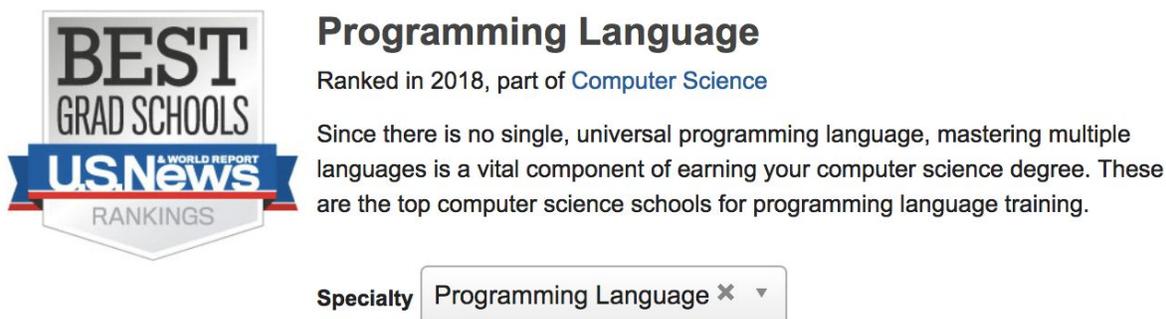

Fig. 1. https://www.usnews.com/best-graduate-schools/top-science-schools/computer-programming-rankings

This lack of understanding of the field suggests that it is highly questionable that *US News and World Report* has the necessary expertise to rank the quality of PhD programs across computer

---

[1] The title and (re-ordered) bullets below are in homage to Dijkstra's famous letter to the CACM [Dijkstra, 1968].

science.  In fact, we know that many rankers often use the wrong data. For example, we have repeatedly seen problems with rankers who only consider journal publications, leaving out conferences, which capture the most influential publications in most areas of Computing.  The consequences are rankings that are completely implausible.  For example, while King Abdulaziz University may be a fine institution, it is unlikely that anyone with any familiarity with Computing-related departments would rank them number 13 in the world, as *US News and World Report* does in its recent ranking of "Best Global Universities." (Fig. 2)

- 
    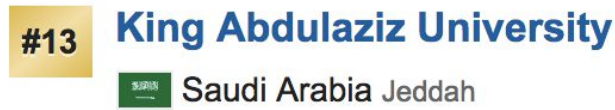
- Fig. 2. https://www.usnews.com/education/best-global-universities/computer-science

A key limitation of a number of rankings, including those produced by *US News and World Report*, is that they depend in whole or in part on reputation surveys.  One problem with reputation is that it is a lagging indicator. When an institution improves, it can take years for its reputation to catch up.  Reputation surveys therefore are inherently "stale".  A more serious problem with reputation surveys is that opinions are often based on subjective assessments with very little basis.

No one is sufficiently knowledgeable about all aspects of Computer Science and all departments to even make an informed *guess* about the broad range of work in an entire department.  In fact, a "mid-rank" department is often the most difficult to assess by reputation because the department may be particularly strong in some sub-areas but weaker in others, i.e., the subjective rating of the department may vary greatly depending on the sub-area of the assessor.

To summarize, rankings matter and will not go away, regardless of their shortcomings.  Commercial rankers today do a poor job of ranking Computer Science departments.  Since we understand our community and what matters, we should take control of the ranking process.

At the very least, we as a community should insist on rankings derived from objective data, whether it be based on publications, citations, honors, funding, or other criteria.  We should ensure that rankings are well-founded, based on meaningful metrics, even if we have diverging perspectives on how best to fold the data into a scalar score or rank. We may still arrive at very different rankings, but we will have a defensible basis for comparisons.

Towards this end, the Computing Research Association (CRA) recently stated that a "methodology [which] makes inferences from the wrong data without transparency" ought to be ignored [Davidson et al 2017].  It has also adopted the following statement about best practices:

> "CRA believes that evaluation methodologies must be data-driven and meet at least the following criteria:

- **G**ood data
    - data have been cleaned and curated
- **O**pen
    - data is available, regarding attributes measured, at least for verification
- **T**ransparent
    - process and methodologies are entirely transparent
- **O**bjective
    - based on measurable attributes"

We call rankings that meet these criteria *GOTO Rankings*. Today, there are at least two GOTO rankings: [csrankings.org](csrankings.org) and [csmetrics.org](csmetrics.org) (both are linked from the site [gotorankings.org](gotorankings.org)). CSrankings is faculty-centric and based on publications at top venues, providing links to faculty home pages, Google Scholar profiles, DBLP pages, and overall publication profiles. It ranks departments by aggregating the full-time tenure-track faculty at each institution. CSmetrics is institution-focused, without regard to department structure or job designations for paper authors. It includes industrial labs and takes citations into account.

These are not the only two reasonable ways to rank departments. One may disagree with the rankings these sites produce, or with their choices of weighting schemes or venue inclusion. But one can clearly understand the basis for each and inspect all or most of the included data. These GOTO rankings are a far cry from the products of most commercial rankers.

**Call to Action**

We call on all Computer Science departments and colleges to boycott reputation-based and non-transparent ranking schemes, including but not limited to US News and World Report:

- Do not fill out their surveys. Deprive these non-GOTO rankings of air, at least for Computer Science.
- Do not promote or publicize the results of such ranking schemes in departmental outlets.
- Discourage University administrators from using reputation based and non-transparent rankings.
- Encourage the use of GOTO Rankings such as CSrankings and CSmetrics as better alternatives.